\documentclass{icrc2009}

\usepackage{graphicx}   % for including figures
\usepackage[caption=false]{caption}    % for captions
\usepackage[font=footnotesize]{subfig} % subfig.sty for a double column floating figure using two subfigures
\usepackage{fixltx2e}
\usepackage{url}

\newcommand{\shorttitle}[1]%
{\markboth{Proceedings of the 31\MakeLowercase{$^{st}$} ICRC, {\L}\'{o}d\'{z} 2009}{#1} }
\newcommand{\etal}{\MakeLowercase{\textit{et al. }}} % "et al."

\begin{document}
\title{The sidereal anisotropy of multi-TeV cosmic rays in an expanding Local Interstellar Cloud}

\author{\IEEEauthorblockN{Y. Mizoguchi\IEEEauthorrefmark{1},
			  K. Munakata\IEEEauthorrefmark{1},
                          M. Takita\IEEEauthorrefmark{2} and
                           J. K\'{o}ta\IEEEauthorrefmark{3}}
                            \\
\IEEEauthorblockA{\IEEEauthorrefmark{1}Physics Department, Shinshu University, Matsumoto, Nagano 390-8621, Japan}
\IEEEauthorblockA{\IEEEauthorrefmark{2}Institute for Cosmic Ray Research, University of Tokyo, Kashiwa 277-8582, Japan}
\IEEEauthorblockA{\IEEEauthorrefmark{3}Lunar and Planetary Laboratory, University of Arizona, Tucson, AZ 87721, USA}
}

% please write the preseter's name and short title (3-4 words maximum)
%    which will appear at the header of the even pages.
\shorttitle{Y. Mizoguchi \etal Sidereal anisotropy in an expanding LIC}
\maketitle

\begin{abstract}
The sidereal anisotropy of galactic cosmic ray (GCR) intensity observed with the Tibet Air Shower (AS) experiment still awaits theoretical interpretation.
The observed global feature of the anisotropy is well reproduced by a superposition of the bi-directional and uni-directional flows (BDF and UDF, respectively) of GCRs.
If the orientation of the deduced BDF represents the orientation of the local interstellar magnetic field (LISMF), as indicated by best-fitting a model to the data, the UDF deviating from the BDF orientation implies a significant contribution from the streaming perpendicular to the LISMF.
This perpendicular streaming is probably due to the drift anisotropy, because the contribution from the perpendicular diffusion is expected to be much smaller than the drift effect.
The large amplitude deduced for the UDF indicates a large spatial gradient of the GCR density.
We suggest that such a density gradient can be expected at the heliosphere sitting close to the boundary of the Local Interstellar Cloud (LIC), if the LIC is expanding.
The spatial distribution of GCR density in the LIC reaches a stationary state because of the balance between the inward cross-field diffusion and the adiabatic cooling due to the expansion.
We derive the steady-state distribution of GCR density in the LIC based on radial transport of GCRs in a spherical LIC expanding at a constant rate.
By comparing the expected gradient with the observation by Tibet experiment, we estimate the perpendicular diffusion coefficient of multi-TeV GCRs in the local interstellar space.
\end{abstract}

 
\section{Introduction}
The galactic cosmic rays (GCR) are high-energy nuclei (mostly protons) produced in our Galaxy. The directional anisotropy of GCR intensity gives us important information on the magnetic structure of the region in space through which GCRs traveled to the Earth. In this paper, we analyze the sidereal anisotropy of $\sim$ 5 TeV GCR intensity observed by the Tibet Air Shower (AS) experiment, which is currently the world's highest precision measurement of GCR intensity in this energy region, utilizing both the high count rate and good angular resolution of the incident direction. The GCR anisotropy in this energy region is free from the solar modulation, while it is still sensitive to the local magnetic field structure with a spatial scale comparable to or larger than the Larmor radius of GCR particles in the Local Interstellar Magnetic Field (LISMF). The sky-map of the directional anisotropy reported by the Tibet AS experiment clearly shows the global feature observed with a statistical significance of more than ten times the statistical error \cite{paper01}.
Fig. 1(a) shows the observed GCR intensity in $5^{\circ}\times5^{\circ}$ pixels in a color-coded format as a function of the right ascension ($\alpha$) on the horizontal axis and the declination ($\delta$) on the vertical axis. The global feature of the anisotropy in Fig. 1(a) is successfully modeled by a combination of the uni-directional flow (UDF) and bi-directional flow (BDF) as
\setlength{\arraycolsep}{0.0em}
  \begin{eqnarray}
 I^{GA}_{n,m}&{}={}&a_{1\perp}\cos \chi_1(n,m:\alpha_1,\delta_1)\nonumber \\
&+&{}a_{1\parallel} \cos \chi_2(n,m:\alpha_2,\delta_2)\nonumber\\
&+&{}a_{2\parallel} \cos^2 \chi_2(n,m:\alpha_2,\delta_2)
  \end{eqnarray}              % 
\setlength{\arraycolsep}{5pt}
\begin{figure}[th]
  \centering
  \includegraphics[width=\linewidth]{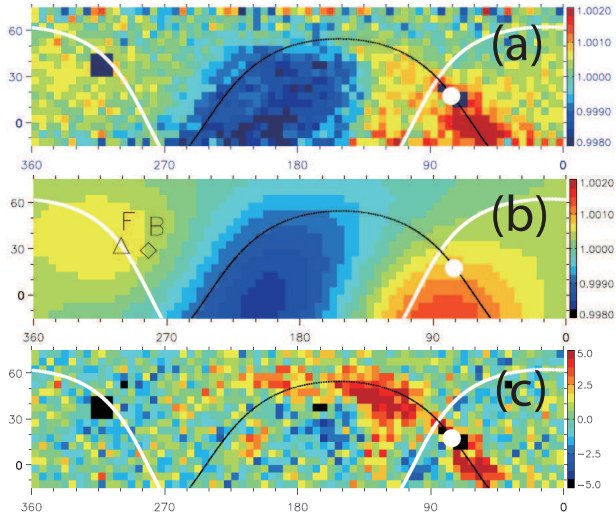}
  \caption{2D-sky maps of the observed and reproduced GCR intensity. Each panel displays the normalized GCR intensity or significance in $5^{\circ}\times5^{\circ}$ pixels in a color-coded format as a function of the right ascension on the horizontal axis and the declination on the vertical axis. In this figure, the average intensity in each declination belt is normalized to unity. These sky-maps cover $360^{\circ}$ of the right ascension but cover only $90^{\circ}$ of the declination due to the event selection criterion limiting zenith angles to $\le 45^{\circ}$ \cite{paper01}. The data in 16 pixels containing the known and possible gamma ray sources are excluded from the best-fit calculation and indicated by black pixels. The white curve indicates the galactic plane, while the black curve displays the HDP plane suggested by Gurnett et al. \cite{paper08}, which is calculated as a plane normal to the orientation of $\alpha=332.1^{\circ}$ and $\delta=35.5^{\circ}$. In each panel, the heliotail direction  ($\alpha=75.9^{\circ}$ and $\delta=17.4^{\circ}$) is indicated by a white solid circle on the HDP plane. Each panel displays, (a): the observed intensity ($I_{n,m}^{obs}$), (b): a component best-fit model anisotropy ($I_{n,m}^{GA}$) reproducing the global anisotropy (GA), (c): the significance of the residual anisotropy remaining after the subtraction of $I_{n,m}^{GA}$ from $I_{n,m}^{obs}$. In the panel (b), the open triangle with an attached character `F' indicates the LISMF orientation by Frisch ($\alpha=300.9^{\circ}$ and $\delta=32.2^{\circ}$) \cite{paper03}, while the open diamond with `B' indicates the orientation of the best-fit BDF (see Table I).
}
  \label{simp_fig}
 \end{figure}
 
\noindent where $I_{n,m}^{GA}$ is the GCR intensity in n-th right ascension and m-th declination pixel representing
the global anisotropy (GA), $a_{1\perp}$ and $a_{1\parallel}$ are amplitudes of UDFs perpendicular and parallel to the BDF, respectively.
$a_{2\parallel}$ is the amplitude of the BDF, $(\alpha_1,\delta_1)$ and $(\alpha_2,\delta_2)$ are respectively right 
ascensions and declinations of the reference axes of the perpendicular UDF and BDF and $\chi_1$ ($\chi_2$) is the angle of the center of ($n, m$) pixel measured from the reference axis of the perpendicular UDF (BDF) \cite{paper01}.
This model anisotropy best-fitting to the data in Fig. 1(a) is displayed in Fig. 1(b), while the residual anisotropy remaining after the subtraction of $I_{n,m}^{GA}$ from the data is shown in Fig. 1(c). It is clear that this model successfully reproduces the global feature in the data. The residual anisotropy in Fig. 1(c) is more local and cannot be reproduced by this simple model for the global feature \cite{paper01}.
The best-fit parameters ($a_{1\perp}, a_{1\parallel}, a_{2\parallel}, \alpha_1, \delta_1, \alpha_2, \delta_2$) are listed in Table I.
\begin{table}[h]
   % \begin{table}[!h]
  \caption{Best-fit parameters in $I_{n,m}^{GA}$. Note that $\delta_2$ in the upper table is derived according to our definition that the reference axes $(\alpha_1,\delta_1)$ and $(\alpha_2,\delta_2)$ are perpendicular to each other \cite{paper02}.}
  \label{table_simple}
  \centering
  \setlength{\tabcolsep}{4.5pt}
  \begin{tabular}{ccccccc}
    \hline \hline 
    $a_{1\perp}(\%)$&$a_{1\parallel}(\%)$&$a_{2\parallel}(\%)$&$\alpha_1(^{\circ})$&$\delta_1(^{\circ})$&$\alpha_2(^{\circ})$&$\delta_2(^{\circ})$\\
    \hline
     0.141 & 0.006 & 0.140 & 37.5 & 37.5 & 102.5 & -28.9 \\
    \hline
  \end{tabular}
 \end{table}
First, we note that the reference axis $(\alpha_2,\delta_2)$ of the BDF is almost parallel to the Galactic plane and very close to the LISMF orientation ($\alpha=300.9^{\circ}$ and $\delta=32.2^{\circ}$) reported by Frisch et al. \cite{paper03}.
This suggests the BDF streaming along the magnetic field.
Second, we note that $a_{1\perp}$ is much larger than $a_{1\parallel}$ indicating the composite UDF almost perpendicular to the BDF.
If the orientation of the deduced BDF represents the orientation of the LISMF, the UDF perpendicular to the BDF orientation implies a contribution from the streaming perpendicular to the LISMF. Adopting the parallel diffusion coefficient of $\kappa_\parallel \sim 10^{29} \hspace{1.5mm} \mbox{cm}^2\mbox{/s}$ by \cite{paper04} together with the Larmor radius of $r_L \sim 0.002 \hspace{1.5mm} \mbox{pc}$ for 5 TeV protons in $3 \hspace{1.5mm} \mu \mbox{G}$ magnetic field, we obtain a Bohm factor $\eta = \lambda_\parallel / r_L \sim 1500$. This large factor implies that the perpendicular diffusion coefficient in the LISMF is much smaller than $\kappa_\parallel$ and the observed UDF perpendicular to the LISMF is mainly due to the drift flow.
The deduced amplitude of the UDF ($a_{1\perp}$) in Table I is as large as $0.14 \hspace{1.5mm} \%$. This large amplitude, if it is arising from the drift flow, requires a small scale structure for producing a large density gradient in the local space. The amplitude of the drift flow, expressed as a vector product ($\mathbf{B} \times \mathbf{\nabla} n$) between the LISMF ($\mathbf{B}$) and the spatial gradient ($\mathbf{\nabla} n$) of GCR density ($n$), is estimated as
 \begin{equation}
 a_{1\perp}=r_L\frac{1}{n}|\mathbf{\nabla}n| \sim r_L \frac{1}{n} \frac{n}{L} =\frac{r_L}{L}.
 \end{equation}
Using the observed $a_{1\perp}=0.141 \hspace{1.5mm} \%$ and $r_L \sim 0.002 \hspace{1.5mm} \mbox{pc}$, we get $L\sim 1.4 \hspace{1.5mm} \mbox{pc}$. This implies that we need to consider the GCR propagation in a very local structure surrounding the heliosphere. As an example for such local structure, we consider the GCR propagation in the Local Interstellar Cloud (LIC). The LIC is an egg-shaped cloud with a volume of $93 \hspace{1.5mm} \mbox{pc}^3$ filled with relatively warm interstellar gas \cite{paper05}. The heliosphere locates inside the LIC close to its boundary. There is another cloud called G cloud approaching to the LIC and the space between two clouds seems to be compressed \cite{paper06}. We suggested that the UDF and BDF may be expected if the GCR density ($n$) is lower inside the LIC than outside, for instance due to the adiabatic expansion of the LIC surrounded by the LISMF. The density gradient is maintained by an equilibrium between the inward diffusion across the LISMF and the adiabatic cooling due to the expansion \cite{paper02}. In this case, the BDF is expected from the parallel diffusion of GCRs into the LIC along the LISMF line connecting the heliosphere on its both ends with the region outside the LIC, where the GCR density is higher. The UDF perpendicular to the LISMF, on the other hand, is expected from the drift anisotropy as described above, if the sufficient density gradient is maintained by the adiabatic expansion of the LIC.
In the next section, we present a simple model for the transport of GCRs into an expanding LIC and discuss the physical parameters required for reproducing the perpendicular UDF observed by the Tibet AS array.

\section{gcr transport into the lic and \newline an estimation of $\kappa_\perp$.}\begin{table*}[h]
  \caption{Diffusion Coefficient and a ratio of coefficient in the interstellar and the heliosphere}
  \label{table}
  \centering
  \begin{tabular}{cccccc}
   \hline \hline
     & $\kappa_\perp (\mbox{cm}^2/\mbox{s})$ & $\kappa_\parallel (\mbox{cm}^2/\mbox{s})$ & $\kappa_T (\mbox{cm}^2/\mbox{s})$ & $\kappa_\perp/\kappa_T$ & $\kappa_\perp/\kappa_\parallel $ \\
   \hline
    The interstellar (5TeV) & $4 \times 10^{24} $ & $10^{29}$ & $6 \times 10^{25}$& 0.07 & $ 4 \times 10^{-5}$\\
    The heliosphere (50GeV) & $3 \times 10^{21}$ & $3 \times 10^{23}$ &$7 \times 10^{21}$ & 0.4 & $10^{-2}$\\
   \hline
  \end{tabular}
%  \end{table}
\end{table*}
We study the transport of GCRs into a spherical LIC. The spherically symmetric distribution of the GCR density in this LIC is governed by the following radial transport equation.
\begin{equation}
 \frac{\partial F}{\partial s}=\kappa_0 \left( \frac{\partial ^2 F}{\partial x^2}+\frac{2}{x} \frac{\partial F}{\partial x} \right) - (2+\gamma )F  \label{eq:a}
\end{equation}
where $F(x,s)$ is the GCR density at a dimensionless distance $x$ from the center of the LIC and at a dimensionless time $s$, while $\gamma=2.7$ is the spectrum index appropriate for high-energy GCRs. We convert the actual radial distance $r$ and time $t$ to $x$ and $s$, respectively, as
\begin{equation}
x=\frac{r}{R(t)}, \hspace{2mm} \mbox{and} \hspace{2mm} s=\log_e{\frac{t}{t_c}}
\end{equation}
where $R(t)$ is the radius of the LIC at time $t$ and $t_c$ is an arbitrary reference time. We assume a self-similar expansion of the LIC with radius $R(t)$ and expansion velocity $V(r,t)$ given as
\begin{equation}
V(r,t)=\frac{r}{t}
\end{equation}
\begin{equation}
R(t)=\frac{R_c t}{t_c}
\end{equation}
with $R_c$  denoting $R(t)$ at $t=t_c$ which we set at  the present time. The first and second terms on the right hand side of eq. (3) describe the inward diffusion, while the third term represents adiabatic cooling due to the expansion. Note that the convection term due to $V$ does not appear in this equation. $\kappa_0$ is a dimensionless parameter depending on the rate of cross-field diffusion and defined by the cross-field diffusion coefficient $\kappa_\perp$ as
\begin{equation}
\kappa_0=\frac{\kappa_\perp}{V_c R(t)}
\end{equation}
where $V_c$ is the expansion velocity of the LIC envelope at $t_c$. We set $R(t)$ to be 2.8 pc according to the reported volume ($93 \hspace{1.5mm} \mbox{pc}^3$) of the LIC \cite{paper05}. The solution of eq. (3) rapidly reaches an equilibrium due to the balance between inward diffusion (causing an increase of $F$) and adiabatic cooling (causing a decrease of $F$). The steady-state solution $F^{\mbox{\small{steady}}}$ is obtained from eq. (3) with the left hand side set equal to zero and given as
\begin{equation}
F^{\mbox{\small{steady}}}(x) = \frac{\sinh(\alpha x)}{x \sinh( \alpha )} \mbox{,} \hspace{4mm} \alpha = \sqrt{\frac{2+ \gamma }{\kappa_0}}\label{eq:b}
\end{equation}
We use this solution to estimate $\kappa_0$ and $\kappa_\perp$. We first calculate the fractional density gradient $100(\partial F/\partial x)/F$ obtained from $F$ in eq. (8) at $x=1$ where the heliosphere is currently located. Fig. 2 displays $100(\partial F/\partial x)/F$ as a function of $\kappa_0$. The fractional density gradient is related to the amplitude $a_{1\perp}$ of the perpendicular UDF as
\begin{equation}
100 \times \Big( \frac{1}{F}\frac{\partial F}{\partial x} \Big)_{x=1}=a_{1\perp}\frac{R(t)}{r_L}
\end{equation}
Substituting the observed $a_{1\perp}=0.141 \hspace{1.5mm} \%$ together with $r_L=0.002 \hspace{1.5mm} \mbox{pc}$ and $R(t)=2.8 \hspace{1.5mm} \mbox{pc}$, we obtain a fractional density gradient of $197.4 \hspace{1.5mm} \%$.
This gradient assumes $\kappa_0=0.47$ (see eq. (8)) which, in turn, gives $\kappa_{\perp}=\kappa_0 V_c R(t)=4 \times 10^{24} \hspace{1.5mm} \mbox{cm}^2\mbox{/s}$, if we assume $V_c \sim 10 \hspace{1.5mm} \mbox{km/s}$ comparable to the speed of the heliosphere relative to the local interstellar medium.
\begin{figure}[!h]
  \centering
  \includegraphics[width=\linewidth]{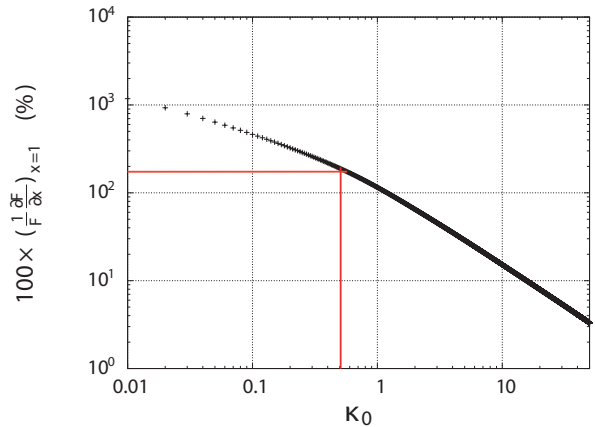}
  \caption{The figure which shows a density gradient of analytic solution on the vertical axis as a function of $\kappa_0$ on the horizontal axis. Red lines indicate a density gradient equal to $197.4\hspace{1.5mm}\%$ and $\kappa_0 = 0.47$.}
  \label{result}
\vspace{10mm}
 \end{figure}
%************
%\begin{table}[!t]
\section{Summary and discussion}
The sidereal anisotropy of the multi-TeV GCR intensity has been observed with an amplitude of $\sim 0.1 \hspace{1.5mm} \%$. If we interprete this amplitude in terms of the UDF due to the parallel diffusion of GCRs in Galaxy, we arrive at
 \begin{equation}
 a_{1\parallel}=\frac{\lambda_\parallel}{L'}.
 \end{equation}
With $a_{1\parallel} \sim 0.1 \hspace{1.5mm} \%$ and $L' \sim$ 100 pc comparable to the scale size of the galactic arm, we obtain $\kappa_\parallel=\lambda_\parallel c/3 \sim 10^{29} \hspace{1.5mm} \mbox{cm}^2\mbox{/s}$ which is comparable to the value reported in \cite{paper04}. This simple interpretation, however, does not hold if the observed UDF is perpendicular to the magnetic field. According to the best-fit analyses of the 2D sky-map of GCR intensity observed by Tibet AS experiment, the UDF with an amplitude of $0.14 \hspace{1.5mm} \%$ is almost perpendicular to the BDF along the LISMF in the galactic plane \cite{paper03}. In this case, we need to consider the GCR propagation in much smaller and more local structures as discussed in I. As an example of such structures, we analyzed the GCR transport into the expanding LIC in this paper, but we do not exclude any other possible local structures with scale sizes comparable to or smaller than the LIC. If the GCR transport in such local structure actually needs to be considered, it also means that we cannot use $\kappa_\parallel$ of \cite{paper04} for interpreting the observed sidereal anisotropy.
Based on our expanding LIC model, we estimate $\kappa_\perp$ as listed in Table II and compare with the value for the heliospheric modulation of GCRs with lower energies \cite{paper07}. In this table, we calculate $\kappa_T$ as
\begin{equation}
\kappa_T=\frac{r_L c}{3}.
\end{equation}
Hence we derive a ratio of $\kappa _{\perp} / \kappa _{T} = 0.07$ for multi TeV GCRs in interstellar space, while this ratio is $\sim 0.4$ for 50 GV GCRs in the heliosphere.
These values are compatible within a factor of six.
On the other hand, the use of $\kappa _{\parallel}$ of \cite{paper04} would result in a 1000 times smaller ratio of $\kappa _{\perp} /\kappa _{\parallel}$ in the interstellar space than in the heliosphere.
This extreme difference tends to suggest that the value of $\kappa _{\parallel} \sim 10^{29} \hspace{1.5mm} \mbox{cm}^2\mbox{/s}$ \cite{paper04} makes it difficult to interpret the observed sidereal variation in terms of GCR transport in very local structures.
 
% see \section{Examples of \LaTeX\  instructions}  \subsection{Tables}

%\vspace{2mm}

\end{document}